\documentclass{article}
\usepackage{spconf,amsmath,graphicx,amssymb,multirow,tabularx,array,booktabs,fancyhdr,hyperref,adjustbox,float}
\usepackage{enumitem}
\usepackage{cite}
\usepackage{hyperref}
\hypersetup{
    colorlinks=true,
    linkcolor=red,
    urlcolor=magenta,
    }

\title{Speaker Adaptation for Quantised End-to-End ASR Models}
\name{Qiuming Zhao\textsuperscript{1}, Guangzhi Sun\textsuperscript{2}, Chao Zhang\textsuperscript{1}, Mingxing Xu\textsuperscript{1}, Thomas Fang Zheng\textsuperscript{1}$^{\ast}$\thanks{$\ast$Correspondence}}
\address{\textsuperscript{1}Tsinghua University, Beijing, China; \textsuperscript{2}University of Cambridge, Cambridge, UK\\
\texttt{\small{zqm23@mails.tsinghua.edu.cn; gs534@cam.ac.uk; \{cz277,xumx,fzheng\}@tsinghua.edu.cn}}}
\begin{document}
\ninept
%\ninept
%
\maketitle
\begin{abstract}
End-to-end models have shown superior performance for automatic speech recognition (ASR) \cite{radford2023robust, gulati2020conformer, zhang2020transformer, kriman2020quartznet}. However, such models are often very large in size and thus challenging to deploy on resource-constrained edge devices. 
While quantisation can reduce model sizes \cite{han2015deep, qian2019binary, leng2018extremely}, it can lead to increased word error rates (WERs). 
Although improved quantisation methods were proposed to address the issue of performance degradation \cite{nguyen2020quantization, ding20224, zhen2022sub, xu2022towards, hernandez2023sharing, yao2020int8}, the fact that quantised models deployed on edge devices often target only on a small group of users is under-explored. 
%For specific applications such as personalised voice assistants or smart door locks, enhancing performance for the individual user takes precedence over generic performance metrics. Consequently, this paper investigates personalisation to compensate for quantisation degradation.
To this end, we propose personalisation for quantised models (P4Q), a novel strategy that uses speaker adaptation (SA) to improve quantised end-to-end ASR models by fitting them to the characteristics of the target speakers. 
%This paper proposes a novel strategy of personalisation for a quantised model (PQM) which performs speaker adaptive training on a quantised end-to-end ASR model. 

In this paper, we study the P4Q strategy based on Whisper and Conformer attention-based encoder-decoder (AED) end-to-end ASR models, which leverages a 4-bit block-wise NormalFloat4 (NF4) approach \cite{dettmers2023qlora} for quantisation and the low-rank adaptation (LoRA) approach \cite{hu2021lora} for SA.
%, which is a parameter-efficient fine-tuning method applied to large models. 
Experimental results on the LibriSpeech and the TED-LIUM 3 corpora show that, with a 7-time reduction in model size and 1\% extra speaker-specific parameters, 15.1\% and 23.3\% relative WER reductions were achieved on quantised Whisper and Conformer AED models respectively, comparing to the full precision models.

%Our main contributions are summarised as follows:
%\begin{itemize}[leftmargin=10pt]
%\setlength\itemsep{0.1em}
%    \item P4Q is the first to use SAT to improve quantised ASR models.
%    \item A pre-training-based LoRA initialisation is proposed for SAT. \item P4Q was evaluated on two commonly used AED ASR baselines across two datasets, yielding consistent improvements.
%\end{itemize}

%Recent advancements in end-to-end automatic speech recognition (ASR) have led to increasingly large models, complicating deployment on resource-constrained devices. While model quantization offers a viable solution, it results in an elevated word error rate (WER). In this paper, we proposed a strategy of personalisation for a quantised model (PQM), which uses personalisation to compensate for quantisation degradation on edge devices. Specifically, PQM adopts a 4-bit NormalFloat Quantisation (NF4) approach for model quantisation and low-rank adaptation (LoRA) for SAT. Experiments on the LibriSpeech and the TED-LIUM 3 corpora show that, with a 7x reduction in model size and 1\% additional speaker-specific parameters, 15.1\% and 23.3\% relative WER reductions were achieved on quantised Whisper and Conformer AED models respectively, comparing to the original full precision models.
\end{abstract}

\section{Methodology}
\label{sec:methodology}

%\subsection{Strategy Overview}
%\label{ssec:strategy overview}

%The PQM strategy consists of three stages. In stage 1, block-wise NF4 quantisation is applied to the model's primary weight parameters. In stage 2, the LoRA, starting with the random initialisation, is pretrained using data from a large set of speakers. In stage 3, we perform speaker-adaptive training on speaker-specific data.
%, during which the entire model is frozen, and only the LoRA parameters corresponding to each speaker are updated.

%\begin{figure*}[t]
%  \centering
%  \includegraphics[width=0.9\linewidth]{overview.pdf}
%  \vspace{-0.5cm}
%  \caption{Overview of the PQM strategy.}
%  \label{fig:Strategy}
  % \vspace{0.2cm}
%\end{figure*}

The P4Q strategy consists of 3 stages: First, block-wise NF4 quantisation is applied to the model's primary weights; then LoRA starting from a random initialisation is pretrained using data from a large set of speakers; at last, SA is applied to the test-time speaker.

\subsection{$k$-bit NormalFloat Quantisation}
\label{ssec:quantisation}

% The block-wise NF4 quantisation is adopted in this paper, which is applied to the weight matrices that are the primary parameters of the model. 
NF4 quantisation equalises the number of values in each bin by determining the quantile of the input matrices through the empirical cumulative normal distribution.
%, as shown in Fig. \ref{fig:NF4}. 
Parameters in a weight matrix are normalised to fit the range of the bins by dividing the maximum absolute value of the matrix. The NF4 quantisation is grounded in the fact that the parameters of a weight matrix generally conform to a normal distribution \cite{dettmers2023qlora}.
%\begin{figure}[t]
%  \centering
%  \includegraphics[width=\linewidth]{NF4.pdf}
%  \vspace{-0.3cm}
%  \caption{\textit{Illustration of the construction of quantiles for NF4 quantisation. It comprises 16 quantisation bins, where the midpoint of each bin represents the quantisation level.}
%  \vspace{-0.3cm}
%}
%  \label{fig:NF4}
%\end{figure}
To mitigate the effect of outliers in weight matrices on maximum absolute value normalisation, block-wise quantisation divides these matrices into blocks. Each block is quantised with its own normalisation factor, confining outliers to their respective blocks. This method enables more fine-grained quantisation by using individual normalisation factors for each block, and results in a smaller quantisation error hence leading to a better performance.

\subsection{LoRA for Speaker Adaptation}
\label{ssec:LoRA for speaker adaptation}

%LoRA is used for SAT to mitigate the loss in performance due to quantisation, which modifies only the model's low-rank subspace parameters, leading to greater efficiency and reduced computation and storage expenses. 
In situations with limited speaker data, full fine-tuning can lead to over-fitting, whereas LoRA can alleviate this issue.
For the pretrained ASR model with weight matrix \( \mathbf{W}_0 \in \mathbb{R}^{d \times k} \), its update is expressed through the following equation, where \( \mathbf{B} \in \mathbb{R}^{d \times r} \) and \( \mathbf{A} \in \mathbb{R}^{r \times k} \), with the rank \( r \ll \min(d, k) \).
\begin{equation}
\mathbf{W}_0 + \Delta \mathbf{W} = \mathbf{W}_0 + \mathbf{B}\mathbf{A}
\end{equation}
% In speaker adaptive training, only the LoRA parameters corresponding to each speaker are updated. In this way, effective adaptation to different speakers can be achieved by updating a minimal set of parameters. Moreover, in cases where the base model is quantised, full-precision LoRA serves to some extent to restore full-precision performance to the base model.
%Given the inherent scarcity of target speaker-specific data, it's common to have more expansive target domain data from other speakers. Such a scenario is usually met in the industry. Therefore, PQM leverages those data to find a better initialisation for LoRA weights via the approach referred to as LoRA pretraining.
%In P4Q, to alleviate the issue of \textit{code-start}, 
Here we propose a LoRA pretraining approach that uses the data of some existing target domain speakers to estimate common initialisation values for the LoRA weights of all test-time speakers.  

% \begin{figure}[t]
%   \centering
%   \includegraphics[scale=0.7]{LoRA.pdf}
%   \vspace{-0.3cm}
%   \caption{\textit{LoRA. Initially, the pretrained weight parameters \( W_0 \) are frozen. For \( A \), random Gaussian initialisation is employed, whereas \( B \) is initialised with zeros.}}
%   \label{fig:LoRA}
%   \vspace{-0.3cm}
% \end{figure}

\section{Experiments}
\label{sec:experiments}
With full NF4 quantisation, the sizes of Whisper and Conformer models were reduced by \textbf{7.25 times} (from 277.8M to 38.3M) and \textbf{6.85 times} (130.9M to 19.1M), and their \%WERs increased from 10.02 to 11.22 and 12.43 to 12.77 respectively. 
%models were reduced by $\sim$7 times, while the WERs for Whisper and a Conformer ASR increased by 1.20\% and 0.34\% respectively. 
%In Table \ref{tab:quantise}, with full NF4 quantisation, the sizes models were reduced by $\sim$7 times, while the WERs for Whisper and a Conformer ASR increased by 1.20\% and 0.34\% respectively. 
%
Table \ref{tab:whisper} shows that compared to the baseline, Whisper-LoRA-pretrain-NF4 achieved a relative WER reduction of 24.2\% on LibriSpeech-SA and 12.8\% on TL3-SA (TED-LIUM 3) sets, showing large improvements over Whisper-FFT-NF4. The same experiments were also performed for Conformer as in Table \ref{tab:conformer}. The Conformer-LoRA-pretrain-NF4 model achieved a WER reduction of 25.3\% compared to the baseline. %These experiments validate the effectiveness of P4Q.

%\newcommand{\tabincell}[2]{\begin{tabular}{@{}#1@{}}#2\end{tabular}}

%\begin{table}[hbtp]
%\caption{\textit{WER on the LibriSpeech-SA using quantised Whisper and Conformer for the linear, convolution, and embedding layers.}}%标题
%\vspace{0.05cm}
%\setlength{\tabcolsep}{3pt}
%\centering%把表居中
%\begin{tabular}{lccc}%四个c代表该表一共四列，内容全部居中
%\toprule
%\textbf{System} & \textbf{Change in \%WER} & \textbf{Change in Size (M)} & \textbf{Reduction}\\
%\midrule
%Whisper & 10.02$\rightarrow$11.22 & 277.8$\rightarrow$38.3 & 7.25 times \\
%\midrule
%Conformer & 12.43$\rightarrow$12.77 & 130.9$\rightarrow$19.1 & 6.85 times \\
%\bottomrule
%\end{tabular}
%\label{tab:quantise}
%\end{table}

 \begin{table}[hbtp]
% \begin{table}[H]
\caption{\textit{\%WERs using quantised Whisper: -baseline means w/ NF4 w/o SA; -FFT is full fine-tuning; -scratch is initialising LoRA weight randomly; -pretrain is full P4Q with LoRA pretraining.}}
% along with subsequent full fine-tuning(Whisper-FFT-FP32), training LoRA from scratch(Whisper-LoRA-scratch), and fine-tuning pretrained LoRA(Whisper-LoRA-pretrain).
\vspace{0.2cm}
\begin{tabular*}{\columnwidth}{@{\extracolsep{\fill}}lcc}
\hline
%\multicolumn{1}{c}{\multirow{2}{*}{\textbf{System}}} & \multicolumn{2}{c}{\textbf{WER(\%)}} \\
%\multicolumn{1}{c}{} & LibriSpeech-SA & TL3-SA \\
\textbf{System} & \textbf{LibriSpeech-SA} & \textbf{TL3-SA} \\
\hline
Whisper-baseline & 11.22 & 7.71 \\
Whisper-FFT-FP32 & 9.59 & 6.85 \\
Whisper-FFT-NF4 & 10.51 & 7.19 \\
Whisper-LoRA-scratch-NF4 & 9.67 & 6.95 \\
Whisper-LoRA-pretrain-NF4 & \textbf{8.51} & \textbf{6.72} \\
\hline
\end{tabular*}
\label{tab:whisper}
% \vspace{0.2cm}
\end{table}

\begin{table}[hbtp]
% \begin{table}[H]
\caption{\textit{\%WERs of Conformer AED: -baseline means w/ NF4 w/o SA. FFT, LoRA; scratch and pretrain are the same as in Table \ref{tab:whisper}}.}
\vspace{0.2cm}
\centering
\begin{tabular}{lc}
\hline
\textbf{System} & \textbf{Librispeech-SA} \\ % 这里不再需要 \multicolumn
\hline
Conformer-baseline & 12.77 \\
Conformer-FFT-FP32 & 10.43 \\
Conformer-FFT-NF4 & 11.99 \\
Conformer-LoRA-scratch-NF4 & 10.25 \\
Conformer-LoRA-pretrain-NF4 & \textbf{9.54} \\
\hline
\end{tabular}
\label{tab:conformer}
\vspace{0.2cm}
\end{table}

\bibliographystyle{IEEEbib}
\bibliography{strings,refs}

\end{document}